# Hybrid Tribo/piezoelectic Electrospun Nanofibers for Energy Harvesting Enhancement in Flexible Electronics

Hao Zhang, Yurong He, Yaofeng Jin, Hui Wang, Wanqi Ye, Lidong Chen, and Kaiyang Zeng*

**ABSTRACT:** Triboelectric nanogenerators (TENGs) and piezoelectric nanogenerators (PENGs) have emerged as promising platforms for harvesting mechanical energy and converting it into electrical energy for powering flexible electronic devices. However, the material selection and structure design of such hybrid nanogenerator, and mechanisms of energy output still remain challenges. In this work, electrospinning is employed for the fabrication of nanofibers, particularly polyvinylidene fluoride (PVDF)-based nanofibers, due to its capability to generate high β-phase contents that effectively increase the piezoelectric performance of the PVDF friction layer, thereby enhancing the overall electrical performance for flexible electronics by merging tribo/piezoelectric power. Furthermore, various concentrations carbon nanotubes (CNT) or graphene nanosheets (GNS) are individually incorporated into the PVDF solution as nanofillers (NF) to enhance the piezoelectric responses of the PVDF-based nanofibers. The introduction of nanofillers is found to not only alter the fiber diameter but also modify the surface roughness of the electrospun nanofibers, and thus, enhancing the triboelectric effect. In addition, the output performance of the fabricated nanogenerator is predominantly governed by the piezoelectric effect rather than triboelectric effect, as the electrical output shows a strong positive correlation with the β-phase content of PVDF-based nanofibers: the highest β-phase content reached to 85.3% and consistently resulted in the optimal energy output of 1.133 watt/m$^2$. Notably, the power density achieved by the prototype device reaches to the level of watt/m$^2$, representing a substantial improvement compared with that of the conventional TENGs or PENGs reported to date, providing expanded opportunities for flexible electronic devices.

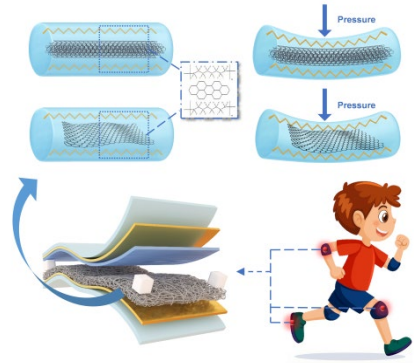

**KEYWORDS:** Electrospinning, Triboelectric nanogenerator, piezoelectric, nanofiller, β-phase PVDF nanofibers, flexible electronics

# 1. INTRODUCTION

The rising global energy consumption has motivated the burgeoning requirement for the replacement of the traditional energy production in 21$^{st}$ century.[1] In this case, a new generation method for energy has turned out to be the emerging case in turns of this global crisis. One possible solution to this severe situation is the application of electroactive materials. Therefore, modern materials such as triboelectric, piezoelectric, and pyroelectric materials have been widely studied due to their outgoing energy harvesting abilities by transferring surrounding mechanical vibrations into electrical power.[2] Compared to the common materials, these electroactive materials can be flexible, durable, portable, and easily fabricated into the desired devices for certain applications, especially for nanogenerators. The idea of triboelectric nanogenerator (TENG) was first proposed by Wang et al.[3-5] Over the past two decades, there are increasing interests in the discovery of possibilities by utilizing TENG since a wide variety of applications have been already covered, including the self-powered sensors,[6] implantable[7] and wearable devices,[8,9] ocean energy harvesting[10] and many more. Traditional ceramic based piezoelectric materials, such as zinc oxide (ZnO),[11-15] barium titanate ($BaTiO_3$),[16-19] lead zirconate titanate ($Pb(Zr,Ti)O_3$ or PZT),[20-23] may show its advantages of possessing high piezoelectric coefficient and dielectric constant.[24,25] However, high brittleness[26] with tiny level of strain limits the integration of wearable devices which mostly need to conform to curved surfaces and withstand significant strain. To overcoming these shortcomings, polymers with piezoelectric properties are superior candidates regarding to the outstanding flexibility and biocompatibility.[27-30] Particularly, PVDF (Polyvinylidene Fluoride) is a promising electroactive semicrystalline polymer with wide frequency responses, relatively high piezoelectric coefficient compared to that of the other polymers, high dipole moment in the polar phases, and high thermal stability.[31-39] Based on the difference in the processing parameters, PVDF is found to have at least five different crystalline phases (α, β, γ, δ, and ε).[40] Among these, β-phase is the most focusing phase since its well-aligned polarization direction that has the highest piezoelectric properties.[41-43] Thus, it gives a fundamental theory that a maximize of the β-phase ratio is desired in the case of the mechanical-electrical energy transfer efficiency, and thus, influencing the following energy harvesting performance. Till now, a wide variety of methods have been studied, including: (1) Mixture of nanofillers;[44] (2) Mechanical stretching;[45] (3) Quenching from the melt;[46] (4) Solvent casting with specific conditions;[47] and (5) Eletrospinning.[48-50] In general, the addition of nanofillers can enhance the formation of the β-phase to some extent by forming strong interactions between the fillers and the PVDF polymer chains. However, this interaction is still not strong enough to suffer the long-term mechanical deformation or sustain repeating strain implementation because of the fundamental issues of inhomogeneity or discontinuity during mixing.[42,51] Apart from the other reported methods, it is also believed that electrospinning is one of the most effective ways to produce high β-phase due to the enormous amount of stretching forces exerted on the PVDF solution during preparation that provided by the high electrostatic field environment.[52] Nevertheless, the issue of β-phase partially depolarization still exists with the further force exertion or even temperature fluctuation.[42] As a consequence, novel approaches to stabilize the β-phase in PVDF nanofibers were introduced by integrating the incorporation of nanofillers and the utilization of electrospinning technique.[52] In this study, we use multi-walled carbon nanotubes (MWCNT) or graphene nanosheets (GNS) as nanofillers respectively with various concentrations because of their excellent electrical conductivity and charge transfer ability, which are favourable for the energy generation efficiency. The nanofiller doped PVDF is followed by one-step electrospinning to fabricate the functionalised PVDF nanofibers, which exhibit significantly enhanced β-phase stability as well as the piezoelectric property. The prepared polymer film composed of functionalised PVDF nanofibers acts as a promoted negative layer, which is an essential component for the energy harvesting efficiency enhancement when assembling the self-powered wearable TENG devices for electronics powering purposes.

## 2. EXPERIMENTS SECTION

### 2.1. Materials.

The raw materials used in this work were all purchased commercially. Polyvinylidene fluoride (Mw = 600,000) powder was procured from Arkema Chemical Co. Ltd. (Shanghai, China). N,N-dimethylformamide (DMF, $C_3H_7NO$, 99.5%) was obtained from Macklin Biochemical Technology Co. Ltd. (China). Acetone (ACE, $C_3H_6O$, 99.5%) was purchased from Yonghua Chemical Co. Ltd. (China). Polyethylene terephthalate (PET, 0.1 mm thick) was purchased from Shenzhen Juli Hardware Plastic Products Co. Ltd. (Shenzhen, China). Furthermore, multi-walled carbon nanotubes (MWCNT, 90%, Tanfeng Graphene Technology Co., Ltd, China) and graphene nanosheets (GNS, 0.5-5 μm, Aladdin Biochemical Technology Co., Ltd, China) were employed as nanofillers (NF). All materials were used as received without further purification.

### 2.2. Preparation of Nanofiller Doped PVDF Film.

PVDF solutions were prepared by dissolving 10 wt% PVDF powder (relative to the total solvent mass) in a solvent system of DMF/ACE (3:2 w/w). Nanofillers (NF) were then incorporated into the solutions at varying concentrations relative to the PVDF mass: 1, 3, 5, and 7 wt% for MWCNT, while 0.75, 1.5, 2.25, and 3 wt% for GNS. The mixtures were magnetically stirred at 500 rpm for 3 h at 60°C to obtain homogeneous dispersions, followed by 1 h of ultrasonication at 60°C to mitigate nanoparticle agglomeration. The resulting suspension was then loaded into a 5 mL syringe fitted with a stainless-steel needle (#22 nozzle). Electrospinning was performed using a high-voltage DC power supply (Dongwen High Voltage Power Co., Ltd，China) with an applied voltage of 15 kV and a feed rate of 1.2 ml $h^{-1}$. Fibers were collected on silicon substrates positioned 12 cm from the needle tip. Ambient conditions were maintained at 26 ± 2°C and 30 ± 5% relative humidity. The nanofiber films were fabricated through continuous electrospinning collection (Figure 1), followed by thermal drying in a convection oven at 60°C for 12 hours. The resultant composite membranes were systematically labelled as PVDF-x wt% CNT (with x = 1.0, 3.0, 5.0, 7.0) and PVDF-y wt% GNS (with y = 0.75, 1.5, 2.25, 3.0) to reflect their respective nanofiller concentrations.

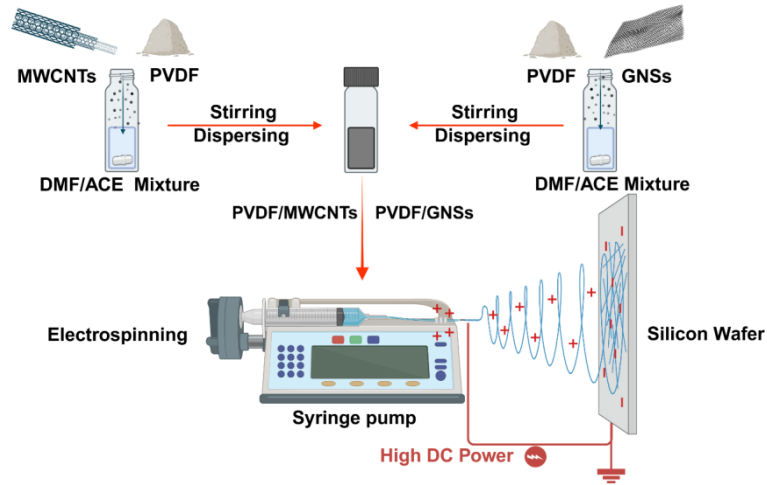

**Figure 1.** Fabrication processes of the electrospun PVDF-based nanofibers.

## 2.3. Characterization and Measurements.

The morphology of the nanofibers was analyzed using a Scanning Electron Microscope (SEM, Quanta 250, Germany) at an accelerating voltage within the range of 10-20 kV. Samples were sputter-coated with a thin gold layer to enhance conductivity before the SEM examination.

Crystallographic analysis was performed on an X-ray Diffractometer (XRD, Rigaku SmartLab, SE, Japan) with the following parameters: Cu-Kα radiation ($\lambda$ = 1.5406 Å), scan range 5° to 50° 2θ, scanning speed 2° 2θ/min, and step size 0.01° 2θ. Chemical functional groups were identified using a Fourier-Transform Infrared Spectroscopy (FTIR, Thermo Fisher Nicolet iS20, Germany) in the Attenuated Total Reflectance (ATR) mode. Spectra were acquired in the wavenumber range of 4000–400 cm$^{-1}$ with a resolution of 4 cm$^{-1}$.

Triboelectric performance of the composite membranes was evaluated using a custom-built triboelectric nanogenerator with a contact-separation mode. The TENG device employed a PVDF-based composite membrane as the negative triboelectric layer and a PET film as the positive triboelectric layer. A motor-controlled stage (PDMS-100, Beijing Jingke Zhichuang Technology Development Co. Ltd., China) regulated the horizontal contact-separation motion at 2 Hz with an applied force of approximately 10 N. The resulting electrical signals, including the open-circuit voltage and short-circuit current, were recorded in real time using an electrometer (Keithley 6514, Tektronix, USA). To determine the maximum output voltage and optimal power output, the voltage across external load resistors (ranging from 1 MΩ to 100 MΩ) were measured using the same electrometer.

## 3. RESULTS AND DISCUSSION

As shown in Figure 2, the morphological differences between untreated PVDF nanofibers and NF-doped PVDF nanofibers are examined using SEM, and the corresponding fiber diameters with deviations are displayed as insects in the top-right corner of each images. All PVDF-based nanofibers are electrospun with relatively uniform size distributions, exhibiting diameters in the range of 550–2050 nm (Table S1, Support Information (SI)) and the diameter changing trend is showed in Figure S1 (SI). Compared with that of the NF-doped PVDF, the pure PVDF nanofibers present smooth surfaces without the presence of particle clusters, as illustrated in Figure 2a. In contrast, the NF-doped PVDF fibers display irregular morphology featuring with small surface protrusions, and the fiber surfaces become noticeably rough. These protrusions are attributed to the embedded nanoparticles, confirming the successful incorporation of nanofillers into the PVDF matrix. In addition, a roughened surface is also beneficial for enhancing the triboelectric generation performance.

For the CNT-doped PVDF nanofibers (0 wt% to 3 wt%), the fiber diameter initially increases upon CNT addition, primarily due to the introduction of CNTs, which facilitates the fiber expansion. With further CNT addition (3 wt% to 5 wt%), the fiber diameter decreases, which can be ascribed to the enhanced electrical conductivity of the spinning solution. The elevated conductivity intensifies the stretching of the electrified jet, as higher charge density leads to the formation of thinner fibers. However, at higher CNT loadings (5 to 7 wt%), the fiber diameters increase again, likely due to CNT agglomeration, which causes localized thickening within the fibers.

For the GNS-doped PVDF nanofibers, a pronounced decrease in fiber diameter is observed when the GNS concentration increased from 0 wt% to 2.25 wt%, yielding the thinnest fibers (~550 nm) at 2.25 wt%. Similar to the case of CNTs, this decrease can be linked to the increased electrical conductivity of the spinning solution; however, the 2D planar structure of GNS provides a more effective conductive network at low loading compared with the 1D unidirectional structure of CNTs. Consequently, enhanced conductivity at lower GNS concentrations strongly promotes the jet stretching. Beyond 3 wt% GNS, particle aggregation becomes significant, leading to an increase in fiber diameter due to the disruption of uniform fiber formation.

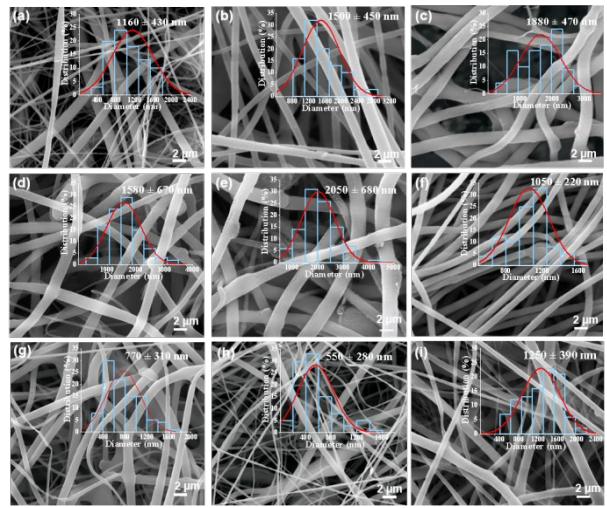

**Figure 2**. SEM images of the: (a) PVDF; (b) PVDF-1 wt%CNT; (c) PVDF-3 wt%CNT; (d) PVDF-5 wt%CNT; (e) PVDF-7 wt%CNT; (f) PVDF-0.75 wt%GNS; (g) PVDF-1.5 wt%GNS; (h) PVDF-2.25 wt%GNS; and (i) PVDF-3 wt%GNS; showing their uniform sizes and average diameters with deviations as insets in the top-right corner of each image.

To evaluate the piezoelectric performance of the nanofiller-modified PVDF fibers, the β-phase serves as a pivotal character that cannot be overlooked. Among the multiple crystalline polymorphs of PVDF, only the β-phase adopts the all-trans (TTTT) chain conformation, which generates a large net dipole moment and thus endows the material with strong piezoelectric activity. Consequently, developing effective strategies to increase the β-phase fraction has emerged as a critical research focus.

To determine the crystallographic form of the nanofiber samples, XRD analysis is performed on blank groups and composite nanofibers with various concentrations of CNT and GNS as shown in Figures 3a,b. The PVDF and its composite nanofibers exhibit primary diffraction peaks at 2θ = 17.7°, 18.4°, and 19.9°, corresponding to the (100), (020), and (110) planes of the α-phase, while the peak at 2θ = 20.2° is corresponding to the (110) plane of the β-phase. It is observed that the gradual introduction of nanoscale fillers effectively suppresses the formation of the α-phase, especially at the peak 2θ =18.4° and promotes the growth of the peak intensity of β-phase. The total degree of crystallinity ($\varphi_{total}$) and the β-phase crystallinity ($\varphi_\beta$) can be calculated using equations (1) and (2), where $\Sigma A_{crys}$ and $\Sigma A_{amr}$ represent the sum of the integral areas of the crystalline and amorphous peaks of PVDF, respectively. Additionally, $\Sigma A_\beta$ represent the sum of the integral areas of the β-phases:

$$\varphi_{total} = \frac{\Sigma A_{crys}}{\Sigma A_{crys} + \Sigma A_{amr}} \times 100\% \qquad \text{(Eq. 1)}$$

$$\varphi_\beta = \frac{\Sigma A_\beta}{\Sigma A_{crys}} \times \varphi_{total} \qquad \text{(Eq. 2)}$$

According to the calculated data, as shown in Figure 3c, with the introduction of nanoscale fillers, the β-phase crystallinity content increased from the lowest 59.7% for pure PVDF to the top of 63.4% for PVDF-5% CNT and 63.2% for PVDF-2.25 wt% GNS in terms of the increasing doping concentrations, respectively. With the excess NF addition, β-phase crystallinity begins to decrease (Table S2, SI). This is attributed to the influence of the nanomaterials on the arrangement of the PVDF molecular chains. At low filling concentrations, NFs promote the alignment of the molecular chains in a fully trans (TTTT) conformation to form highly polar β-phase crystals. However, excess NFs embedded in the PVDF matrix disrupt the orientation and chain alignment and thus lowering its crystallinity. Therefore, only a proper amount of NFs (5 wt% for CNT, 2.25 wt% for GNS) that yielded a homogeneous PVDF matrix and resulted in the highest β-phase crystallinity.

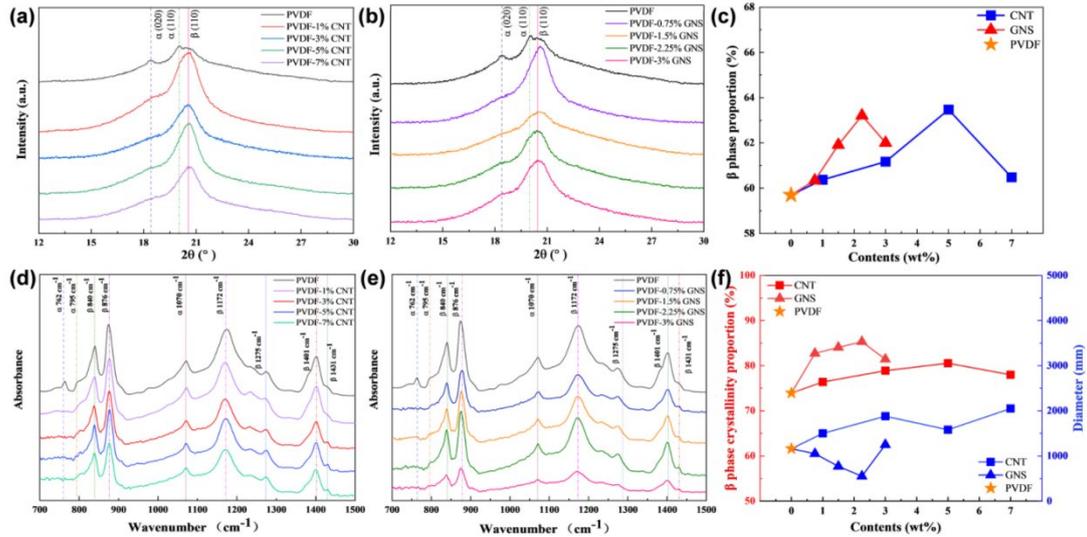

**Figure 3.** XRD pattern of (a) PVDF and CNT-doped PVDF with various CNT concentrations; (b) PVDF and GNS-doped PVDF with various GNS concentrations; (c) PVDF and CNT/GNS-doped PVDF β phase crystallinity content with various individual dopant concentrations; (d) FTIR spectra of PVDF and CNT-doped PVDF with various CNT concentrations; (e) FTIR spectra of PVDF and GNS-doped PVDF with various GNS concentrations; and (f) The correlation between β phase proportion and fiber diameter in terms of dopant type and its various concentrations.

The crystallinity and its correlation with NF concentration are further examined using FTIR analysis. As shown in the FTIR spectra (Figures 3d,e), to identify the presence of the α- and β-phases, the characteristic absorption bands of PVDF and NF-doped PVDF fibers are analyzed within the range of 70 - 1500 cm$^{-1}$. The bands at 762, 795, 1070, and 1275 cm$^{-1}$ are attributed to the α phase, whereas those at 840, 876,

1172, 1275, 1401, and 1431 cm$^{-1}$ are characteristic of the β phase.[44] Notably, to compare the NF-doped PVDF fibers with that of the pristine PVDF, the intensity of the α-phase peak at 762 cm$^{-1}$ exhibits a substantial reduction, and in some cases nearly disappears. As previously discussed, this peak corresponds to the α-phase, and its attenuation indicates that the introduction of nanofillers effectively modifies the intermolecular chain alignment, thereby facilitating the α-to-β phase transition. A similar trend is also observed at 1431 cm$^{-1}$, where the enhanced β-phase band further confirms the nanofiller-induced promotion for the β-phase formation in electrospun PVDF fibers.

Regarding the effect of CNT incorporation, pure PVDF fibers or samples with low CNT loading (1 wt%) exhibit a broad and weak peak at 1431 cm$^{-1}$, which is often negligible. This suggests that low filler concentrations are insufficient to induce noticeable β-phase formation. However, with increasing CNT content, the intensity of the β-phase peak at 1431 cm$^{-1}$ progressively increases, indicating a clear enhancement in β-phase crystallization. This behavior is attributed to the fact that nanofillers commonly possess polar functional groups or surface charge that interact with the -CH$_2$- and -CF$_2$- dipoles of PVDF chains. Such interactions facilitate the alignment of molecular chains into the all-trans (β) conformation during crystallization. Specifically, CNTs exhibit high surface electron density, which can induce electrostatic polarization and promote chain stretching, further assisting β-phase formation.

A similar trend is observed for the GNS-doped PVDF fibers. The characteristic α-phase peak at 762 cm$^{-1}$ diminishes and eventually disappears as the GNS concentration increase, while the β-phase peak at 1431 cm$^{-1}$ shows a pronounced rise in its intensity. As GNS also possesses high surface electron density, its incorporation effectively drives PVDF chains toward the all-trans β orientation. Consequently, a predictable enhancement in β-phase content can be obtained.

To quantitatively determine the β-phase fraction ($F_{(\beta)}$), the β-phase peaks are deconvoluted and integrated, and the β-phase content is calculated using Equation 3, where $A_i$ represents the peak area at wavenumber $i$ cm$^{-1}$. The relationship between the β phase fraction and fiber diameter is plotted in Figure 3f.

$$F_{(\beta)} = \frac{A_{840}+A_{876}+A_{1172}+A_{1275}+A_{1401}+A_{1431}}{A_{762}+A_{795}+A_{840}+A_{876}+A_{1070}+A_{1172}+A_{1275}+A_{1401}+A_{1431}} \qquad \text{(Eq. 3)}$$

The FTIR results exhibit an overall trend consistent with the XRD analysis regarding the evolution of β-phase content. For CNT-doped PVDF fibers, the β-phase content increases with CNT loading (0-3 wt%), which can be attributed to electrostatic polarization induced by nanofillers at low concentrations. In this regime, nanofiller-polymer interactions dominate the nucleation of the β-phase, while the contribution from the electrical stretching force remains relatively limited. However, with further CNT addition, an inverse relationship between β-phase fraction and fiber diameter becomes evident, whereby thinner fibers correspond to higher β-phase content. This phenomenon is mainly associated with the increased electrical conductivity of the spinning solution, which enhances elongational forces during electrospinning. The resulting mechanical stress exerted on the jet is recognized as a major driving force promoting β-phase formation.

In the case of GNS-doped PVDF fibers, the two-dimensional planar structure of GNS provides higher effective conductivity within the solution, enabling the rapid formation of continuous conductive networks. Consequently, an inverse correlation between β-phase content and fiber diameter appears even at the initial stage of GNS incorporation, indicating that electrical stretching plays a dominant role from the outset. Overall, these observations suggest that the electrical stretching force becomes the primary mechanism governing β-phase enhancement at higher nanofiller concentrations, surpassing the nucleation effect associated with nanofiller-matrix interactions.

Regarding the absolute β-phase content, all nanofiller-doped PVDF fibers show significantly higher β-phase fractions relative to that of the pure PVDF fibers. The maximum β-phase content reaches 85.3% at 2.25 wt% GNS, while CNT-doped PVDF attains a peak value of 80.5% at 5 wt% CNT. At the same doping concentration (3 wt%), GNS-doped PVDF exhibits a slightly higher β-phase content compared to its CNT counterpart. Nevertheless, excessive nanofiller loading leads to a sharp decline in β-phase, which can be attributed to agglomeration and the resulting disruption of polymer chain alignment. As illustrated in the figure, both systems show a critical concentration, beyond which over-doping induces detrimental effects on chain configuration. Collectively, CNT and GNS serve as effective agents for promoting formation of the β-phase in the electrospun PVDF, but only within an optimal concentration window.

The type and concentration of nanofillers influence both the morphological and piezoelectric characteristics of PVDF nanofibers and also exert a pronounced effect on their charge storage and energy harvesting capability. To evaluate the performance of the mechanical-to-electrical conversion, a cyclic compression system applying a periodic load of 10 N at 2 Hz is employed. The as-prepared films by electrospun are trimmed into 3×3 cm squares, while the effective contact area of the TENG device is fixed at 3 cm² to ensure reliable comparison across all samples (Figure S2). The resulting open-circuit voltage ($V_{oc}$) and short-circuit current ($I_{sc}$) are then recorded to determine the optimal output performance of each composite film.

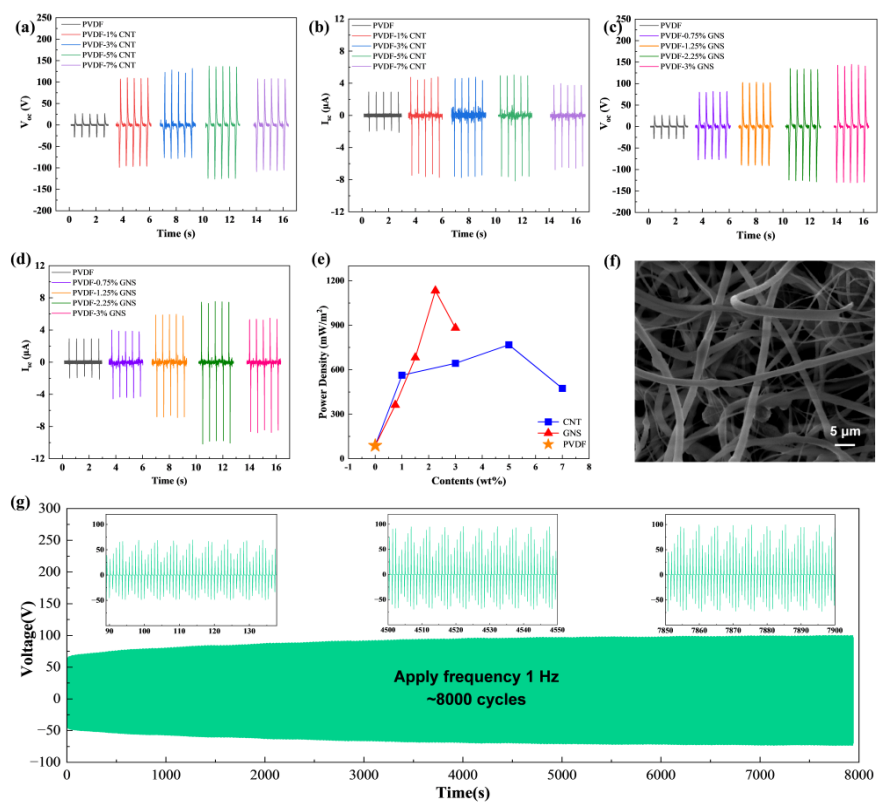

**Figure 4.** (a) The open-circuit voltage and (b) short-circuit current of PVDF and CNT-doped PVDF with various CNT concentrations; (c) The open-circuit voltage and (d) short-circuit current of PVDF and GNS-doped PVDF with various GNS concentrations; (e) Power density of PVDF and CNT/GNS-doped PVDF with various individual dopant concentrations; (f) SEM image of hybrid nanogenerator after 8000 cycles compression; and (g) Output voltage durability of the hybrid nanogenerator (partially enlarged view of the initial, middle, and final stages).

As shown in Figures 4a,b, the pristine PVDF nanofibers produce the lowest electrical outputs, exhibiting only 27 V and 2.9 μA in $V_{oc}$ and $I_{sc}$, respectively, owing to the limited intrinsic β-phase content. For CNT-doped PVDF nanofibers, upon increasing the CNT content, both $V_{oc}$ and $I_{sc}$ increase significantly, reaching peak values of 137.4 V and 5 μA at 5 wt%. Beyond this concentration, however, a marked deterioration in performance is observed, which is attributed to CNT agglomeration that suppresses dipole orientation and disrupts the electrospinning jet stability.

In contrast, GNS-doped PVDF fibers exhibit a more rapid and pronounced improvement. The open-circuit voltage (Figure 4c) increases steadily with GNS concentration and reached a plateau after 2.25 wt%, consistent with the superior conductivity and planar morphology of GNS, which promotes more effective electric-field-induced chain alignment during electrospinning. Meanwhile, the short-circuit current increases sharply, as indicated in Figure 4d, peaking at 7.55 μA before declining to approximately 5.5 μA at higher concentrations due to particle aggregation and reduced chain mobility.

A more quantitative detailed comparison is obtained by calculating the power density (Figure S3, SI). As shown in Figure 4e, for trend comparison, CNT-doped PVDF fibers demonstrate a monotonic rise in power density up to 757.9 mW/m² at 5 wt%, after which a slight decrease can be observed. A similar trend is noted for GNS-doped fibers, with a distinct maximum of 1132.8 mW/m² at 2.25 wt%, which is nearly 13 times higher than that of the pristine PVDF (88.2 mW/m²). Remarkably, this watt-level power density exceeds the majority of previously reported PVDF-based triboelectric or piezoelectric nanogenerators, indicating a significant advancement in output performance. These results highlight the effectiveness of nanofiller-mediated chain orientation and crystalline phase regulation in enhancing the energy conversion of electrospun PVDF.

Furthermore, the variation in power density exhibits an almost identical trend to the evolution of β-phase content as discussed earlier. Such strong correlation suggests that the dominant mechanism governing the electrical output in this TENG system is the piezoelectric effect rather than the triboelectric effect. Therefore, the optimized nanofiller concentrations for electrospinning: 5 wt% for CNT and 2.25 wt% for GNS, represent the most favorable conditions for maximizing β-phase content, dipole alignment, and ultimately, energy harvesting performance.

The operational stability of the fabricated hybrid nanogenerator is systematically evaluated to assess its long-term durability and reliability. A continuous cyclic compression test is conducted under practical working conditions, where a periodic compressive force of 10 N at a frequency of 1 Hz is applied for a total duration of 8000 s, corresponding to 8000 loading cycles.

The long-term performance stability is comprehensively verified through both structural and electrical characterizations. Specifically, the morphological integrity of the electrospun nanofibers after the cyclic compression test is examined, together with continuous monitoring of the output voltage and current signals throughout the entire testing period. As shown in Figure 4f, the nanofibers maintain their structural integrity without observable fracture, delamination, or fatigue-induced damage even after undergoing 8000 repeated mechanical impact cycles. It is worth noting that the applied compressive force of 10 N is considerably higher than the loading conditions commonly reported in related studies, which typically range from 3 N to 5 N, thereby imposing a more stringent mechanical challenge on the device. The ability of the nanofiber-based system to withstand such a high compressive stress without structural degradation further demonstrates the excellent mechanical robustness and structural resilience of the composite fibers, highlighting their suitability for long-term operation under practical and demanding working conditions. Notably, an evident charge accumulation process is observed during the initial ~1000s of the contact–separation cycles, during which the output voltage gradually increased from approximately 60 V to a maximum value of around 100 V (Figure 4g). This behavior can be attributed to the progressive buildup of surface charges and the stabilization of interfacial polarization under repeated mechanical stimulation. Beyond this initial charging stage, the electrical output signals exhibit negligible degradation over prolonged operation, indicating highly stable and repeatable voltage and current responses. The combination of preserved nanofiber morphology and consistent electrical output strongly confirms the excellent mechanical robustness and electrical stability of the hybrid nanogenerator. These results indicate that the fabricated device is well-suited for long-term operation, highlighting its promising potential for practical applications in wearable electronics and self-powered sensing systems.

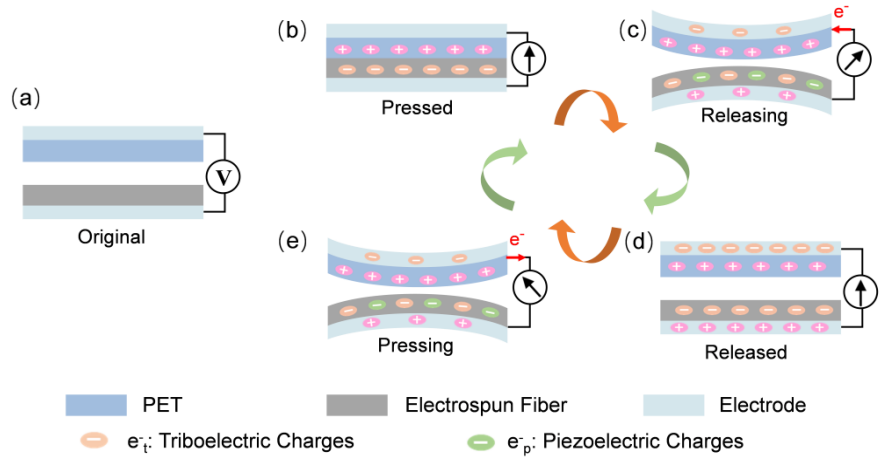

**Figure 5.** Hybrid triboelectric and piezoelectric working mechanisms of the device under different states: (a) original; (b) pressed; (c) releasing; (d) released; and (e) pressing.

Figure 5 schematically shows the fabricated nanogenerator working processes and illustrates the basic working mechanism under vertical periodically compression. The hybrid nanogenerator is generally composed of two friction layers which are PET and electrospun PVDF fiber (Figure 5a). In comparison with PET, the electrospun PVDF layer is more favourable in obtaining electrons and thus acting as the negative friction layer. Accordingly, PET is prone to losing electrons and generating positive charges on the friction surface, and only triboelectric effect can be applied on this surface. However, when PVDF is used as a triboelectric layer, both triboelectric and piezoelectric effects contribute to the electrical output. As illustrated in figures 5b-c, upon contact with PET layer, PVDF tends to gain electrons due to its high electronegativity, leaving negative triboelectric charges ($e^-_t$) on its surface after separation. Under external pressure, the PVDF fibers are deformed and intrinsic dipoles within PVDF undergo reorientation, leading to enhanced piezoelectric polarization and the generation of bound charges ($e^-_p$) on its surfaces. The combined surface tribo/piezoelectric charges and pressure-induced polarization modify the electric potential difference between the electrodes, forcing free electrons in the external circuit to flow in opposite directions during the pressing and releasing processes (Figures 5c,e). When fully released, the two friction layers are separated and two electrodes remain stable with equal but opposite charges due to the electrostatic induction (Figure 5d). Consequently, a coupled triboelectric-piezoelectric mechanism is established, enabling efficient energy conversion under mechanical stimuli.

To demonstrate the practical applicability of the assembled hybrid nanogenerator, a battery-free electronic stopwatch is employed as a proof-of-concept load. As shown in Figure 6a, also in the supplementary video (Video S1, SI), the electrical energy harvested from the device is first stored in a capacitor and subsequently utilized to power the stopwatch through continuous mechanical compression. Under an applied load of 10 N at a frequency of 2 Hz, the capacitor is sufficiently charged within approximately 35 s to successfully activate the stopwatch. Furthermore, the as-designed hybrid nanogenerator is capable of directly lighting a series of light-emitting diodes (LEDs) through simple manual pressing, without the need for additional external circuits, as demonstrated in Figures 6b-d, as well as in the supplementary video (Video S2, SI). The number of LEDs that can be simultaneously illuminated serves as an intuitive yet effective indicator of the power generation capability of the nanogenerator. In this work, the hybrid nanogenerators based on pristine PVDF, PVDF–5 wt%CNT, and PVDF–2.25 wt%GNS are able to power approximately 274, 515, and 635 LEDs, respectively. This pronounced increase in the number of illuminated LEDs clearly reflects the enhanced electrical output induced by nanofiller incorporation. Moreover, the observed trend is in excellent agreement with the previously discussed power density output results, further validating the effectiveness of CNT and GNS in boosting the energy harvesting performance of the PVDF-based hybrid nanogenerators. This result clearly indicates that the output power generated by the hybrid nanogenerator can drive low-power electronic devices without the need for external batteries, highlighting its strong potential for self-powered wearable and portable electronic applications.

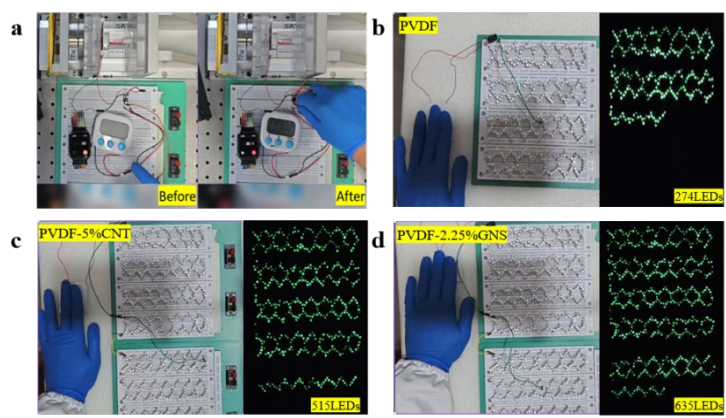

**Figure 6.** (a) Charging of a portable timer clock; (b) Powering of LEDs using PVDF based hybrid nanogenerator; (c) Powering of LEDs using PVDF-5 wt%CNT based hybrid nanogenerator; and (d) Powering of LEDs using PVDF-2.25 wt%GNS based hybrid nanogenerator.

## 4. CONCLUSIONS

In summary, a hybrid TENG/PENG system incorporating specially doped PVDF nanofibers, fabricated via electrospinning, is successfully developed with markedly enhanced power-generation capability. Both pristine and nanofiller modified PVDF nanofibers are systematically characterized to identify the optimal nanofiller type and its corresponding loading concentration. Among all tested samples, PVDF-2.25 wt%GNS exhibit the highest tribo/piezoelectric output, achieving a power density of 1132.8 mW/m$^2$, which is approximately 13 times higher than that of the untreated PVDF nanofibers. Furthermore, the output performance shows a strong correlation with the β phase content of the electrospun nanofibers, indicating that the electrical output of the hybrid device is predominantly governed by the piezoelectric effect rather than the triboelectric effect. The hybrid TENG/PENG demonstrated in this work broadens the applicability of self-powered flexible electronics that are previously constrained by high power requirements, offering expanded possibilities for smart wearable technologies by harvesting low-frequency mechanical energy from human activities.

# ASSOCIATED CONTENT

## Supporting Information

The Supporting Information is available free of charge at：

    Nanofiber Diameter Determination; Statistical Analysis; The size difference in nanofiber diameter with corresponding deviations between PVDF and two NF-doped PVDF nanofibers with various concentrations; The diameter trend between PVDF and two NF-doped PVDF nanofibers with various concentrations; The differences in β phase crystallinity proportions between PVDF and two NF-doped PVDF nanofibers with various concentrations; Figure of cyclic compression system and the assembled TENG/PENG; The differences in power density between PVDF and two NF-doped PVDF nanofibers with various concentrations; Video of powering stopwatch and LEDs (MP4)

# AUTHOR INFORMATION

## Corresponding Author

  **Kaiyang Zeng** - National University of Singapore (Suzhou) Research Institute, 377 Lin Quan Street, Suzhou Industrial Park, Suzhou 215123, China；Department of Mechanical Engineering, National University of Singapore, 9, Engineering Drive 1, Singapore 117542, Singapore; Email: mpezk@nus.edu.sg

## Authors

  **Hao Zhang** - National University of Singapore (Suzhou) Research Institute, 377 Lin Quan Street, Suzhou Industrial Park, Suzhou 215123, China; Email: hao.zhang@nusri.cn

  **Yurong He** - Department of Mechanical Engineering, National University of Singapore, 9, Engineering Drive 1, Singapore 117542, Singapore

**Yaofeng Jin** - National University of Singapore (Suzhou) Research Institute, 377 Lin Quan Street, Suzhou Industrial Park, Suzhou 215123, China

**Hui Wang** - National University of Singapore (Suzhou) Research Institute, 377 Lin Quan Street, Suzhou Industrial Park, Suzhou 215123, China

**Wanqi Ye** - National University of Singapore (Suzhou) Research Institute, 377 Lin Quan Street, Suzhou Industrial Park, Suzhou 215123, China

**Lidong Chen** - State Key Laboratory of High Performance Ceramics, Shanghai Institute of Ceramics, Chinese Academy of Sciences, Shanghai 200050, P.R.China

## Author Contributions

L.C. and K.Z. designed the research. H.Z. drafted the original manuscript, performed the characterization and analyzed data. Y.H. conducted experimental work and visualization experiments. Y.J. conducted the power output test. H.W. and W.Y. provided the laboratory support. K. Z. revised and finalized the manuscript.

## Notes

The authors declare no conflict of interest.

# ACKNOWLEDGMENTS

This work was supported by Science and Technology Project of Jiangsu Province, grant number: BZ2022056. K.Z also thanks the financial support from Ministry of Education (MOE) Singapore through the National University of Singapore (NUS) under the Academic Research Funds (Nos. A-0009122-01-00).